\documentclass[twocolumn,showpacs,floatfix]{revtex4}%
\usepackage{graphicx}%
\usepackage{amsmath}%
\usepackage{amsfonts}%
\usepackage{amssymb}
\usepackage{bm}

\def\s{{\sigma}}
\def\e{{\epsilon}}
\def\k{{ {\bm k} }}
\def\p{{ {\bm p} }}
\def\q{{ {\bm q} }}
\def\Q{{ {\bm Q} }}

\def\0{{ {\bm 0} }}
\def\w{{\omega}}
\def\a{{\alpha}}

\allowdisplaybreaks[4]

\begin{document}
\title{
Theoretical Prediction of the Nematic Orbital-Ordered State \\
in Ti-Oxypnictide Superconductor BaTi$_2$(As,Sb)$_2$O
}
\author{
Hironori \textsc{Nakaoka}$^{1}$, 
Youichi \textsc{Yamakawa}$^{1}$, and 
Hiroshi \textsc{Kontani}$^{1}$
}

\date{\today }

\begin{abstract}
The electronic nematic state without magnetization
emerges in various strongly correlated metals
such as Fe-based and cuprate superconductors.
To understand this universal phenomenon,
we focus on the nematic state 
in Ti-oxypnictide BaTi$_2$(As,Sb)$_2$O,
which is expressed as the three-dimensional 10-orbital Hubbard model.
The antiferromagnetic fluctuations are caused by
the Fermi surface nesting.
Interestingly, we find the spin-fluctuation-driven 
orbital order due to the strong orbital-spin interference, 
which is described by the Aslamazov-Larkin vertex correction (AL-VC).
The predicted intra-unit-cell nematic orbital order is
consistent with the recent experimental reports on BaTi$_2$(As,Sb)$_2$O.
Thus, the spin-fluctuation-driven orbital order due to the AL-VC mechanism 
is expected to be universal
in various two- and three-dimensional multiorbital metals.

\end{abstract}

\address{
$^1$ Department of Physics, Nagoya University,
Furo-cho, Nagoya 464-8602, Japan. 
}
 
\pacs{71.45.Lr,74.25.Dw,74.70.-b}
%74.25.Dw Superconductivity phase diagrams
%74.70.-b Superconducting materials other than cuprates
%71.45.Lr Charge-density-wave systems

\sloppy

\maketitle

%%%%%%%%%%%%%%%%%%
\section{Introduction}
\label{sec:intro}
%%%%%%%%%%%%%%%%%%

Various interesting symmetry breaking phenomena
associated with the charge, orbital, and spin degrees of freedom
emerge in strongly correlated electron systems.
Among them, the rotational symmetry breaking,
so called the nematic transition,
has attracted increasing attention, after the discovery of the 
nematic order in Fe-based and cuprate superconductors.
In Fe-based superconductors,
both the spin-nematic order 
\cite{Fernandes,DHLee,Chubukov,QSi}
and the orbital order 
\cite{Kruger,PP,WKu,Onari-SCVC,Onari-SCVCS,Text-SCVC,JP}
are considered as the origin of the nematic order.
In cuprates, the $p$-orbital order is a promising candidate
for the nematic  \cite{Yamakawa-CDW,Tsuchiizu-CDW},
whereas other promising scenarios had been proposed so far
\cite{DHLee-cup,Kivelson-cup,Chubukov-cup,Sachdev-cup}.
The nematic transitions in these superconductors
cannot be understood within the random-phase-approximation (RPA)
based on the Hubbard models,
so it is demanded to develop the microscopic theory beyond the 
mean-field-level approximations.
%so we should develop the microscopic theory beyond the 
%mean-field-level approximations.

%To understand the origin of the electronic nematic order
%and its formation mechanism,
To achieve the fundamental understanding of the electronic nematic states,
we focus on the nematic phase in the Ti-oxypnictide superconductors
\cite{YajimaJPSJ2012,YajimaJPSJ2013,Zhai2013,Wang2010}.
No magnetic order appears in the nematic phase \cite{NMR,Rohr},
similarly to FeSe.
%The non-magnetic nematic phase realized in these Ti-oxypnictides
The nematic order in Ti-oxypnictides is driven by the 
electron-interaction since the orthorhombic lattice deformation
$(a-b)/(a+b)$ is just $\sim0.1$\% \cite{Frandsen2014},
which is even smaller than that in Fe-based superconductors.
In contrast to these systems, the lattice distortions in 
Jahn-Teller systems (like Mn-oxides) reach a few \%.
The nematic order in  BaTi$_2$As$_2$O is ascribed to
the intra-unit-cell charge-density-wave (CDW) with $d$-wave symmetry
since no superlattice was found by the 
electron diffraction studies \cite{Frandsen2014,Nozaki}.
This result is analogous to the ``$d$-symmetry CDW'' in under-doped cuprates
in Refs. \cite{Yamakawa-CDW,Tsuchiizu-CDW,STM-Fujita-cup}.
Such intra-unit-cell CDW in Ti-oxypnictide is unable to be 
explained by the electron-phonon mechanism \cite{Subedi}.
Thus, the study of Ti-oxypnictides should serve to 
understand the origin of the nematicity due to the electron-interaction.
%Coulomb interaction.

Interestingly, the superconducting phase (with $T_{\rm c}\sim5$K)
is realized near the nematic phase in various Ti-oxypnictides,
indicating the importance of the nematic fluctuations on the superconductivity.
In addition, strong antiferromagnetic (AFM) spin fluctuations appear
near the nematic phase in BaTi$_2$Sb$_2$O \cite{NMR},
analogously to the Fe-based and cuprate superconductors.
Therefore, Ti-oxypnictides would give us great hints 
to understand the close interplay between the nematicity,
magnetism, and superconductivity,
which is a central issue in Fe-based and cuprate superconductors.

% punch line
In this paper, we study the origin of the non-magnetic nematic order
in Ti-oxypnictides
based on the realistic Hubbard model for BaTi$_2$(As,Sb)$_2$O.
Due to the Fermi surface (FS) nesting, the AFM fluctuations develop 
at $\Q_s=(\pi,0,\pi)$ and $(0,\pi,\pi)$, consistently with the 
previous theoretical studies \cite{Yu,Singh} 
and the NMR study \cite{NMR}.
Remarkably, we find that the strong orbital fluctuations at $\q=(0,0,0)$
are induced by the Aslamazov-Larkin vertex correction (AL-VC),
which is neglected in the RPA.
We predict the formation of the spin-fluctuation-driven 
intra-unit-cell orbital order in BaTi$_2$(As,Sb)$_2$O.
Thus, the nematic orbital order due to the AL-VC
is realized in Ti-oxypnictides, 
similarly to Fe-based and cuprate superconductors.

The AL-VC represents the orbital-spin interplay,
which is intuitively 
understood in terms of the strong-coupling picture $U\gg W_{\rm band}$
as we explained in Ref. \cite{Onari-SCVCS}:
In Fe-based superconductors, the ferro-orbital order 
$n_{xz}\gg n_{yz}$ gives rise to 
the strong anisotropy in the nearest-neighbor exchange interaction,
$J^{(1)}_x\ne J^{(1)}_y$,
and therefore the stripe AFM order with is induced.
Thus, the orbital-order/fluctuations and magnetic-order/fluctuations
simultaneously emerges.
Such Kugel-Khomskii-type orbital-spin interplay 
is explained by the AL-VC in the weak-coupling picture.

%%%%%%%%%%%%%%%%%%
\section{Model Hamiltonians}
\label{sec:model}
%%%%%%%%%%%%%%%%%%
% model

%%%%%%%%%%%%%%%%%%%%%%%%%%%%%%%%%%%%%%%%
\begin{figure}[htbp]
\includegraphics[width=0.9\linewidth]{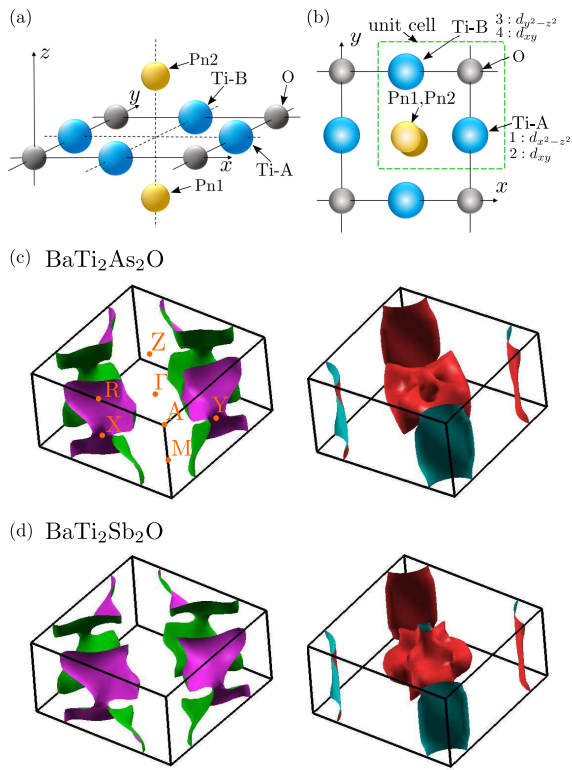}
\caption{
(a) Crystal structure of Ti$_2$Pn$_2$O-layer and 
(b) unit cell with two Ti-ions and two Pn-ions.
(c) Three-dimensional FSs for BaTi$_2$As$_2$O and 
(d) FSs for BaTi$_2$Sb$_2$O 
obtained by the band calculation.
}
\label{fig:3DFS}
\end{figure}
%%%%%%%%%%%%%%%%%%%%%%%%%%%%%%%%%%%%%%%%

Figures \ref{fig:3DFS} (a) and (b) 
show the metallic Ti$_2$Pn$_2$O-layer in Ti-oxypnictides
(Pn=As, Sb).
The unit cell contains two Ti-sites and two Pn-sites.
The bandstructure near the Fermi level is mainly composed of
($d_{x^2-z^2}$, $d_{xy}$)-orbitals on Ti-A and
($d_{y^2-z^2}$, $d_{xy}$)-orbitals on Ti-B,
which are respectively denoted as (1, 2) and (3, 4) hereafter.
Here, we perform the band calculation of BaTi$_2$Pn$_2$O
using the WIEN2K software, 
In Fig. \ref{fig:3DFS}, we show the FSs for (c) BaTi$_2$As$_2$O
and the FSs for (d) BaTi$_2$Sb$_2$O
given by the WIEN2K software.
The crystal structures are respectively given in 
Ref. \cite{Wang2010} and Ref. \cite{YajimaJPSJ2012}.
In both compounds, the FSs are composed by 
the hole-like cylinder FSs around X and Y points,
the electron-like cylinder FS around M point,
and three-dimensional FSs around $\Gamma$ point.
The shape of the three cylinder FSs are very similar in both compounds.
Although the shape of the three-dimensional FSs is different 
between these compounds,
these FSs play a minor role on the orbital fluctuation mechanism.
%The FSs of the present model for another Ti-oxypnictide
%BaTi$_2$(As$_{0.5}$Sb$_{0.5}$)$_2$O
%are given by the average between the FSs in
%Figs. \ref{fig:3DFS} (a) and (b).

\begin{figure}[!htb]
\includegraphics[width=.9\linewidth]{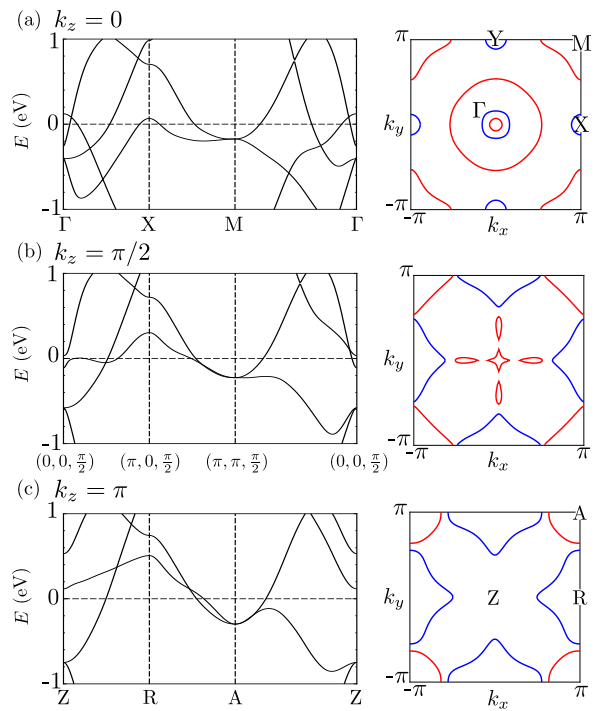}
\caption{
(color online)
(a)-(c) Bandstructure and FSs in the $k_z=0$, $k_z=\pi/2$, 
and $k_z=\pi$ planes.
The red and blue lines correspond to electron-like FSs 
and hole-like FSs, respectively.
}
\label{fig:FS}
\end{figure}

Next, we derive three-dimensional 10-orbital tight-binding model
for BaTi$_2$Pn$_2$O, with the four $d$-orbitals 
(orbital $1\sim4$) and six $p$-orbitals of Pn1,2 
(orbital $5\sim10$), using the WANNIER90 software.
The tight-binding hopping parameters for BaTi$_2$(As$_{1-x}$Sb$_{x}$)$_2$O
is approximately given by the interpolation between the parameters 
for BaTi$_2$As$_2$O and the parameters for BaTi$_2$Sb$_2$O.
In this paper, we present the numerical study for 
BaTi$_2$(As$_{0.5}$Sb$_{0.5}$)$_2$O: 
We verified that the numerical results are essentially unchanged by $x$.
The electron number per unit cell 
(two Ti ions and two Pn ions) is $14.0$.
%The $d$-electron number per Ti site is $1$,
%and $p$-electron number per Pn sites is $6$.
The bandstructure and the FSs in the 
$k_z=0$, $k_z=\pi/2$, and $k_z=\pi$ planes are shown in 
Figs. \ref{fig:FS} (a)-(c), respectively.
The red (blue) lines represents the electron-like (hole-like) FSs.
The two hole-like cylinder FSs around X,Y points and the 
one electron-like cylinder FS around M point give the 
dominant density-of-states (DOS) at the Fermi level.
The nesting between these FSs gives the 
spin fluctuations at $Q_s=(\pi,0,\pi)$ and $(0,\pi,\pi)$.
In addition, complex three-dimensional FSs exist
around $\Gamma$ point.
%Hereafter, the unit of energy is eV.

%%%%%%%%%%%%%%%%%%%%%%%%%%%%%%%%%
%%%%%%%%%%%%%%%%%%%%%%%%%%%%%%%%%

The multiorbital Coulomb interaction term is given as
\begin{eqnarray}
&&\!\!\!\!\!\!\!\!\!\!
H_{\rm M}^U=
\frac12 \sum_{i,l,\s\ne\s'} U n_{i,l\s}n_{i,l,\s'}
+\frac12 \sum_{i,l\ne m,\s,\s'} \left\{ U'n_{i,l\s}n_{i,m,\s'} \right.
\nonumber \\
&&\left. +J_{m,l}c_{i,m\s}^\dagger c_{i,l\s}
(c_{i,l\s'}^\dagger c_{i,m\s'}+c_{i,m\s'}^\dagger c_{i,l\s'} \delta_{\s,-\s'}) \right\},
\end{eqnarray}
where $U$ and $U'$ are the 
intra-orbital and inter-orbital Coulomb interactions,
and $J$ is the Hund's interactions for $d$-electrons.
Hereafter, we assume the relation $U=U'+2J$ and $J>0$,
and fix the ratio $J/U=1/9$ except in Fig. \ref{fig:SCVC2} (d):
We verified that similar numerical results are obtained for $J=1/8$.
In the case of Fe-based superconductors,
$J/U$ ranges  from 0.0945 (in FeSe) to 0.134 (in LaFeAsO)
according to the detailed and exhaustive first-principles study 
in Ref. \cite{Arita}.

%%%%%%%%%%%%%%%%%%
\section{Theoretical Analysis}
\label{sec:theory}
%%%%%%%%%%%%%%%%%%

Based on the obtained model, we calculate the spin and orbital 
susceptibilities.
The bare susceptibility ($U=0$) is
$\chi_{l,l';m,m'}^0(q)=-T\sum_{k}G_{l,m}(k+q)G_{m',l'}(k)$,
%
%\begin{eqnarray}
%\chi_{l,l';m,m'}^0(q)=-T\sum_{k}G_{l,m}(k+q)G_{m',l'}(k),
%\end{eqnarray}
%
where $l,m$ are the orbital indices, $k=(\k,\e_n)$ and $q=(\q,\w_l)$;
$\e_n=\pi(2n+1)T$ ($\w_l=2\pi lT$) is the fermion (boson)
Matsubara frequency.
${\hat G}(k)=(i\e_n+\mu-{\hat h}(\k))^{-1}$
is the Green function, where ${\hat h}(\k)$ is the 
kinetic term in the orbital basis.
The charge (spin) susceptibility is
\begin{eqnarray}
{\hat \chi}^{c(s)}(q)=
(1-{\hat \Phi}^{c(s)}(q){\hat \Gamma}^{c(s)})^{-1}{\hat \Phi}^{c(s)}(q),
\label{eqn:chi}
\end{eqnarray}
where 
${\hat \Phi}^{c(s)}(q)= {\hat \chi}^0(q)+ {\hat X}^{c(s)}(q)$
is the charge (spin) irreducible susceptibility, and 
${\hat X}^{c(s)}(q)$ is VC for the charge (spin) channel:
${\hat X}^{c(s)}(q)$ gives the important orbital-spin interference 
although it is dropped in the RPA \cite{Onari-SCVC}.
${\hat \Gamma}^{c(s)}$ is the $d$-orbital bare Coulomb interaction
for the charge (spin) channel, composed of the on-site Coulomb interactions
$U$, $U'$ and $J$ \cite{Onari-SCVC}.
In a single-orbital model, ${\hat \Gamma}^{c(s)}$ is simply given as
${\Gamma}^{s}=U$ and ${\Gamma}^{c}=-U$.

The charge (spin) susceptibility diverges when the 
charge (spin) stoner factor $\a_{C(S)}$,
which is given by the maximum eigenvalue of 
${\hat \Phi}^{c(s)}(\q,0){\hat \Gamma}^{c(s)}$, reaches unity.
%In the RPA, we drop any VCs in the susceptibility.
%In this case, both  $\a_S$ and $\a_C$ increase from zero
With increasing $U$ under the condition $J/U=1/9$,
both $\a_C$ and $\a_S$ increase monotonically,
and the orbital order (magnetic order) occurs when $\a_C=1$ ($\a_S=1$).
In the RPA, in which the susceptibility is given as
${\hat \chi}^{c(s),{\rm RPA}}(q)=
(1-{\hat \chi}^0(q){\hat \Gamma}^{c(s)})^{-1}{\hat \chi}^0(q)$, 
the relation $\a_S>\a_C$ is always satisfied for a positive $J$
\cite{Onari-SCVC}.
Therefore, for $J/U\sim O(10^{-1})$, ${\hat \chi}^{c,{\rm RPA}}(\q)$ remains small 
even when ${\hat \chi}^{s,{\rm RPA}}(\Q_s)$ diverges.

%%%%%%%%%%%%%%%%%%
\subsection{RPA analysis for the spin susceptibility}
\label{sec:rpa}
%%%%%%%%%%%%%%%%%%

\begin{figure}[!htb]
\includegraphics[width=.9\linewidth]{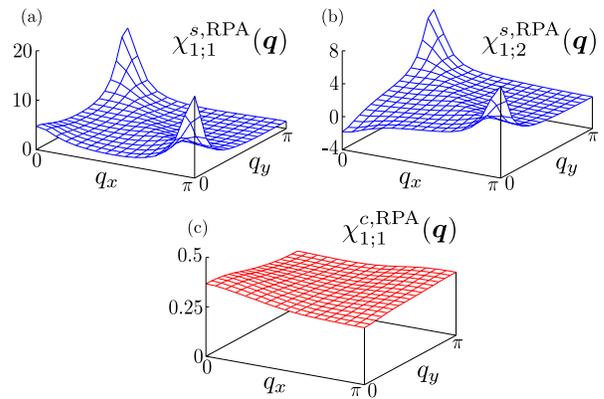}
\caption{
(color online)
(a),(b) RPA spin susceptibilities
 $\chi^{s,{\rm RPA}}_{1;1}(\q)$ and $\chi^{s,{\rm RPA}}_{1;2}(\q)$
%$\chi^{s,{\rm RPA}}_{2,2;2,2}(\q)$ 
at $q_z=\pi$.
Note that $\chi^{s,{\rm RPA}}_{2;2}(\Q_s)\approx6$.
(c) RPA charge susceptibility
$\chi^{c,{\rm RPA}}_{1;1}(\q)$.
We put $U=2.06$ eV and $J/U=1/9$.
}
\label{fig:RPA}
\end{figure}

First, we explain the RPA results obtained by using
$32\times32\times8$ $\bm k$-meshes and $256$ Matsubara frequencies.
We fix the temperature at $T=50$ meV.
Figure \ref{fig:RPA} shows the spin susceptibility in the RPA,
$\chi^{s,{\rm RPA}}_{l;m}(\q)\equiv \chi^{s,{\rm RPA}}_{l,l;m,m}(\q)$, for
(a) $l=m=1$ and (b) $(l,m)=(1,2)$ at $q_z=\pi$
in the case of $U=2.06$ eV and $J/U=1/9$ ($\a_S=0.98$).
They have sharp peaks at $\q=(\pi,0)$ and $(0,\pi)$.
(Note that $\chi^{s,{\rm RPA}}_{l,l';m,m'}(\q)$ is very small for 
$l,l'\le2$ and $m,m'\ge3$.)
The strong spin fluctuations are actually observed in BaTi$_2$Sb$_2$O 
by NMR measurement above the structure transition temperature $T_S$
\cite{NMR}.
However, the charge susceptibility remains very small in the RPA,
as we show  $\chi^{c,{\rm RPA}}_{1;1}(\q)$ in Fig. \ref{fig:RPA} (c).
Thus, the experimental nematic order cannot be explained by the RPA.

%%%%%%%%%%%%%%%%%%%%%%%%%%%%%%%%%
%%%%%%%%%%%%%%%%%%%%%%%%%%%%%%%%%

%%%%%%%%%%%%%%%%%%
\subsection{Analysis of the Aslamazov-Larkin Vertex Correction
for the orbital susceptibility}
\label{sec:al}
%%%%%%%%%%%%%%%%%%

In the next stage, we study the charge susceptibility 
by taking the AL-VC into account, and derive the 
intra-unit-cell orbital order.
In various two-dimensional multiorbital metals
such as Fe-based and cuprate superconductors,
the spin-fluctuation-driven orbital order and fluctuations are 
realized due to the large AL-VC for the charge channel
\cite{Onari-SCVC,Kontani-Raman,Tsuchiizu-Ru1}.
In this mechanism,
very weak spin fluctuations give rise to the orbital order
when the ratio $J/U$ is small, 
%so the orbital order is realized by very weak spin fluctuations in FeSe
as observed in FeSe with ${\bar J}/{\bar U}\approx 0.1$
\cite{FeSe-Yamakawa,Ishida-NMR,Dresden-NMR}.
However, it is highly nontrivial whether the AL-VC is important or not
in three-dimensional systems like Ti-oxypnictides.
%and it is a big challenge to calculate the AL-VC
%in three-dimensional models.
In this paper, we calculate the AL-VC 
in the three-dimensional model for the first time.
We neglect the AL-VC for the spin channel and the Maki-Thompson VCs
since they are negligible in various models 
\cite{Onari-SCVC,FeSe-Yamakawa,Yamakawa-CDW,Tsuchiizu-CDW}.
%For the same reason, we also neglect the Maki-Thompson VCs.
We drop the feedback effect from ${\hat \chi}^c$ to 
${\hat X}^c$ since its smallness has been verified in the present model,
similarly to the case of the $d$-$p$ Hubbard model 
for cuprates \cite{Yamakawa-CDW}.
The expression for the AL-VC is given
in Appendix \ref{sec:AP-A}.

%%%%%%%%%%%%%%%%%%%%%%%%%%%%%%%%%
\begin{figure}[!htb]
\includegraphics[width=.9\linewidth]{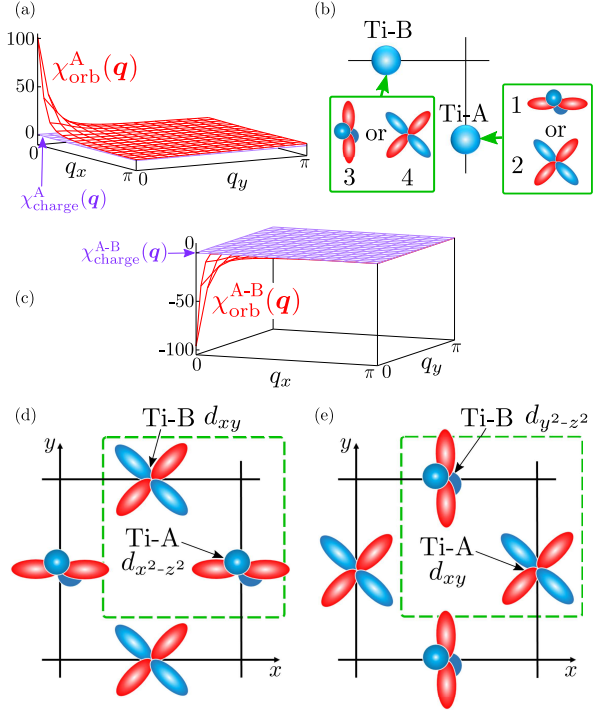}
\caption{
(color online)
(a) $\chi^{\rm A}_{\rm orb}(\q)$ and $\chi^{\rm A}_{\rm charge}(\q)$ at $q_z=0$
%Orbital and charge susceptibilities at Ti-A site
obtained by including the AL-VC.
We put $U=2.06$ eV and $J/U=1/9$.
(b) Schematic intra-site orbital polarization.
(c) $\chi^{\rm A\mbox{-}B}_{\rm orb}(\q)$
and $\chi^{\rm A\mbox{-}B}_{\rm charge}(\q)$ at $q_z=0$.
%Inter-site orbital susceptibility 
%Strong ferro-orbital fluctuations are realized by the AL-VC.
(d),(e) Schematic intra-unit-cell nematic orbital order
$(\Delta n_{x^2-z^2}^{\rm Ti\mbox{-}A},\Delta n_{xy}^{\rm Ti\mbox{-}A},
\Delta n_{y^2-z^2}^{\rm Ti\mbox{-}B},\Delta n_{xy}^{\rm Ti\mbox{-}B}) 
\propto \pm(1,-1,-1,1)$,
which triggers the orthorhombic structural transition.
}
\label{fig:SCVC}
\end{figure}
%%%%%%%%%%%%%%%%%%%%%%%%%%%%%%%%%

We present the numerical results 
%of the orbital and charge susceptibilities
obtained by including the AL-VC.
The Stoner factors are $(\a_C,\a_S)=(0.99,0.98)$ 
for $U=2.06$ eV and $J/U=1/9$ at $T=50$ meV.
Figure \ref{fig:SCVC} (a) shows the susceptibility 
of the orbital polarization at Ti-A site,
\begin{eqnarray}
\chi^{\rm A}_{\rm orb}(\q)&\equiv& \sum_{l,m=1}^2\chi^c_{l;m}(\q)\cdot (-1)^{l+m}
\nonumber \\
&=&\chi^c_{1;1}(\q)+\chi^c_{2;2}(\q)-2\chi^c_{1;2}(\q) ,
\end{eqnarray}
at $q_z=0$.
%That is, 
%$\chi^{\rm A}_{\rm orb}(\q)=\chi^c_{1;1}(\q)+\chi^c_{2;2}(\q)-2\chi^c_{1;2}(\q)$.
%$\chi^{\rm A}_{\rm orb}(\q)\equiv \sum_{l,m=1}^2\chi^c_{l;m}(\q)\cdot (-1)^{l+m}
%=\chi^c_{1;1}(\q)+\chi^c_{2;2}(\q)-2\chi^c_{1;2}(\q)$ at $q_z=0$.
We see that the strong ferro-orbital fluctuations appear due to the AL-VC,
which are absent in the RPA result in Fig. \ref{fig:RPA} (c).
In contrast, the susceptibility of the charge density at Ti-A site,
$\chi^{\rm A}_{\rm charge}(\q)\equiv \sum_{l,m=1}^{2}\chi^c_{l;m}(\q)
=\chi^c_{1;1}(\q)+\chi^c_{2;2}(\q)+2\chi^c_{1;2}(\q)$,
remains small as shown in Fig. \ref{fig:SCVC} (a).
%In addition, $\chi^c_{2;m}({\bm 0})\approx 23$ for $m=2,3$,
%and $\chi^c_{2;4}({\bm 0})\approx -21$.
These results means the emergence of the intra-site orbital polarization,
$\Delta n_1 \Delta n_2<0$ at Ti-A and $\Delta n_3 \Delta n_4<0$ at Ti-B,
as schematically shown in Fig. \ref{fig:SCVC} (b).
Here, $\Delta n_l$ is the modulation of the electron density on orbital $l$.
%in the ordered state.
%The orbital polarization is schematically shown in 

In addition, the inter-site orbital susceptibility between Ti-A and Ti-B,
\begin{eqnarray}
\chi^{\rm A\mbox{-}B}_{\rm orb}(\q)&\equiv& \sum_{l,m=1}^{2}
\chi^c_{l;m+2}(\q)\cdot (-1)^{l+m}
\nonumber \\
&=&\chi^c_{1;3}(\q)+\chi^c_{2;4}(\q)-\chi^c_{1;4}(\q)-\chi^c_{2;3}(\q),
\end{eqnarray}
has large negative peak at $\q={\bm 0}$
as we show in Fig. \ref{fig:SCVC} (c).
In contrast, the inter-site charge susceptibility 
$\chi^{\rm A\mbox{-}B}_{\rm charge}(\q)\equiv \sum_{l,m=1}^{2}\chi^c_{l;m+2}(\q)$
is not enhanced at all.
%Therefore, $\Delta n_1 \Delta n_3<0$ and $\Delta n_2 \Delta n_4<0$ 
%in the ordered state.
These results mean that the orbital polarization 
in the ordered state,
$\Delta {\bm n}\equiv (\Delta n_1,\Delta n_2,\Delta n_3,\Delta n_4)$,
is roughly proportional to $\pm(1,-1,-1,1)$.
%From Fig. \ref{fig:SCVC} (a)-(d),
%we find that the symmetry breaking associated with the orbital polarization
%$\Delta n \equiv n_{x^2-z^2}^{\rm Ti1}-n_{y^2-z^2}^{\rm Ti2}>0$ and
%$\Delta n' \equiv n_{xy}^{\rm Ti1}-n_{xy}^{\rm Ti2}<0$, or 
%$\Delta n<0$ and $\Delta n'>0$.
%(In the tetragonal phase,
%$\Delta n=\Delta n'=0$ by symmetry, and the relation 
%$n_{x^2-z^2}^{\rm Ti1} \approx n_{xy}^{\rm Ti1}$ is satisfied.)

%Now, we obtain the orbital polarization
%numerically:
More properly, the orbital polarization $\Delta {\bm n}$
%$(\Delta n_1,\Delta n_2,\Delta n_3,\Delta n_4)$ 
is proportional to the form factor $\bm f$, which is given by the 
eigenvector of ${\hat \Phi}^{c}(q){\hat \Gamma}^{c}$ for the 
largest eigenvalue $\alpha_{c}$,
as we discussed in Ref. \cite{Yamakawa-CDW}.
%The form factor is proportional to the charge density modulation 
%$(\Delta n_{x^2-z^2}^{\rm Ti1},\Delta n_{xy}^{\rm Ti1},
%\Delta n_{y^2-z^2}^{\rm Ti2},\Delta n_{xy}^{\rm Ti2})$ 
%in the orbital order state \cite{Yamakawa-CDW}. 
The obtained form factor for Fig. \ref{fig:SCVC} 
is $\bm f=\pm(1.06,-0.94,-1.06,0.94)$. 
%is $\bm f=\pm(0.53,-0.47,-0.53,0.47)$. 
Thus, the predicted orbital patter are shown in
Fig. \ref{fig:SCVC} (d) or (e):
When the electron densities for orbitals 1 and 4 in Fig. \ref{fig:SCVC} (d) 
increase, the densities for other orbitals in (e) decrease.
The predicted {\it intra-unit-cell orbital order} 
is consistent with the absence of the superlattice in 
BaTi$_{2}$(As,Sb)$_{2}$O in the nematic phase 
reported by the electron diffraction study 
\cite{Frandsen2014,Nozaki}.

The charge pattern in Fig. \ref{fig:SCVC} (d),(e)
may be safely called orbital-selective charge order,
since it is the spontaneous symmetry breaking among degenerate 
orbitals on different sites.
(Two orbitals on the same Ti-ion are non-degenerated.)
However, we call this charge pattern the orbital order for simplicity, 
since the net charge at Ti-A and that at Ti-B are almost equivalent.

%is consistent with key experimental facts.
%Actually, it is supported by the absence of the superlattice in 
%BaTi$_{2}$(As,Sb)$_{2}$O reported by the recent 
%electron diffraction study 
%\cite{Frandsen2014,Nozaki}.

%%%%%%%%%%%%%%%%%%
\subsection{Explanation for the intra-unit-cell orbital order due to the AL-VC}
\label{sec:reason}
%%%%%%%%%%%%%%%%%%

\begin{figure}[!htb]
\includegraphics[width=.9\linewidth]{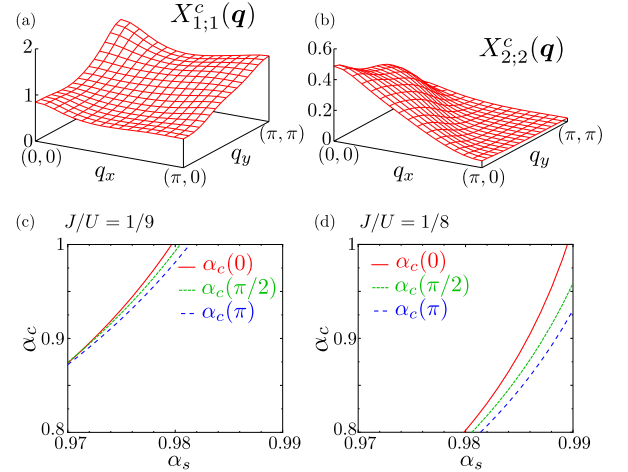}
\caption{
(color online)
(a),(b) Obtained AL-VCs $X_{1;1}^c(\q)$ and $X_{2;2}^c(\q)$ at $\q_z=0$
for $J/U=1/9$.
(c) $\a_C(q_z)$ for $J/U=1/9$ and  
(d) $\a_C(q_z)$ for $J/U=1/8$ 
as functions of $\a_S$.
% in the cases of $J/U=1/9$ and $J/U=1/8$, respectively.
}
\label{fig:SCVC2}
\end{figure}

Here, we verify numerically that the intra- and inter-orbital fluctuations
in Fig. \ref{fig:SCVC} originates from the {\it diagonal elements of the AL-VC};
$X_{l;l}^c({\bm 0})$ with $l=1\sim4$.
For $U=2.06$ eV,
the diagonal elements of the AL-VCs are shown in Figs. \ref{fig:SCVC2} 
(a) and (b), in which
$X_{1;1}^c({\bm 0})=0.84$ and $X_{2;2}^c({\bm 0})=0.49$ respectively.
Then, the irreducible susceptibilities are
$\Phi_{1;1}({\bm 0})=1.21$ and $\Phi_{2;2}({\bm 0})=0.75$.
By taking only the diagonal AL-VCs into account in ${\hat \Phi}^c$, 
strong orbital fluctuations 
%similar to Figs. \ref{fig:SCVC} (a) and (d) are obtained:
with the form factor 
${\bm f}=\pm(1.04,-0.96,-1.04,0.96)$ appears at $U\approx2.0$ eV.
%at $U=2.06$ eV.

%%%%%%%%%%%%%%%%%%%%%%%%%%%%%%%%%
%%%%%%%%%%%%%%%%%%%%%%%%%%%%%%%%%

%${\bm f}=\pm(0.52,-0.48,-0.52,0.48)$.
%irrespective that the magnitude of $X_{1,1;2,2}^c({\bm 0})$ is not small.

Next, we present a mathematical explanation why 
the strong orbital fluctuations with the form factor
${\bm f}\propto (1,-1,-1,1)$ are obtained, which corresponds to the 
intra-unit-cell orbital order.
% shown in Figs. \ref{fig:SCVC} (d) and (e).
To understand this nontrivial result, we put 
$\Phi_{l,l}({\bm 0})=\Phi$ for $l=1,3$ and
$\Phi_{l,l}({\bm 0})=(1-x)\cdot \Phi$ for $l=2,4$,
and other elements are zero for simplicity.
We also put $U=U'$ and $J=0$ for simplicity.
Under this simplification, the largest eigenvalue of 
${\hat \Phi}^{c}(q){\hat \Gamma}^{c}$ is doubly degenerate,
and the corresponding form factors are
${\bm f}_1\propto (y,-1,-y,1)$ and ${\bm f}_2\propto(y,-1,y,-1)$,
where $y\approx 1+x/4$ for $|x|\ll1$,
%where $y=(\sqrt{1-x+x^2/16} -x/4)/(1-x) \approx 1+x/4$ for $|x|\ll1$,
as we explain in Appendix \ref{sec:AP-B}.
This degeneracy of the form factor 
is lifted by the small inter-site components of 
$\Phi_{l;m}({\bm 0})$ with $l\le2$ and $m\ge3$.
In the present model, the form factor ${\bm f}_1$ is selected
mainly due to the negative $X^c_{1;3}({\bm 0})$,
as explained in Appendix \ref{sec:AP-B}.
Therefore, the intra-unit-cell orbital order with
${\bm f}_1\approx(1,-1,-1,1)$
is stably obtained without tuning model parameters.

%Thus, the form factor of the intra-unit-cell orbital order 
%would be insensitive to the model parameters.

In the present theory,
the charge Stoner factor $\a_C$ is enlarged by the AL-VC,
and the AL-VC increases near the magnetic critical point.
Figures \ref{fig:SCVC2} (c) and (d) shows the charge Stoner factor
for a fixed $q_z$, $\a_C(q_z)$, as function of $\a_S$
for $J/U=1/9$ and $1/8$, respectively.
In both cases,
$\a_C(q_z)$ at $q_z=0$ reaches unity for the smallest $\a_S$,
meaning that the orbital order at $\q=(0,0,0)$ is realized.
The orbital order should trigger the 
experimental orthorhombic structure transition at $T=T_S$.

%%%%%%%%%%%%%%%%%%
\subsection{Pseudo-gap formation in the orbital-ordered state}
\label{sec:pg}
%%%%%%%%%%%%%%%%%%

Below, we discuss the electronic states in the ordered state below $T_S$,
by introducing the orbital-dependent potential energy $\Delta E_l$ ($l=1\sim4$).
The potential energy for the intra-unit-cell orbital order
in Fig. \ref{fig:SCVC} (d) or (e) is 
$\Delta {\bm E}_{\rm orbital} \equiv (\Delta E,-\Delta E,-\Delta E,\Delta E)$.
%$\Delta {\bm E}_{\rm orbital} \equiv (1,-1,-1,1)\times\Delta E$.
%$\Delta {\bm E} \equiv (\Delta E,-\Delta E,-\Delta E,\Delta E)$.
%\equiv (\Delta E_1,\Delta E_2,\Delta E_3,\Delta E_4)=
In addition, we also discuss the intra-unit-cell charge order
$\Delta {\bm E}_{\rm charge} \equiv (\Delta E,\Delta E,-\Delta E,-\Delta E)$.
%$\Delta {\bm E}_{\rm charge} \equiv (1,1,-1,-1)\times\Delta E$.
%$\Delta {\bm E}' \equiv (\Delta E,\Delta E,-\Delta E,-\Delta E)$.
This possibility had been discussed in Ref. \cite{Frandsen2014}.
We note that the orbital-ordered state in Fe-based superconductors
($n_{xz}\ne n_{yz}$)
had been explained theoretically, by developing the 
self-consistent vertex correction (SC-VC) theory 
for the orbital-ordered state \cite{Onari-FeSe}, and the 
experimental orbital polarization $E_{yz}-E_{xz}$ 
is $50\sim60$ meV.

%in the orthorhombic state 
%In Fe-based superconductors, 
%the orbital polarization $E_{yz}-E_{xz}$ in the orthorhombic state 
%is $\sim60$ meV in BaFe$_2$As$_2$ \cite{ARPES-Shen}, and 
%$\sim50$ meV in FeSe \cite{ARPES-Shimojima} at X point. 

\begin{figure}[!htb]
\includegraphics[width=.99\linewidth]{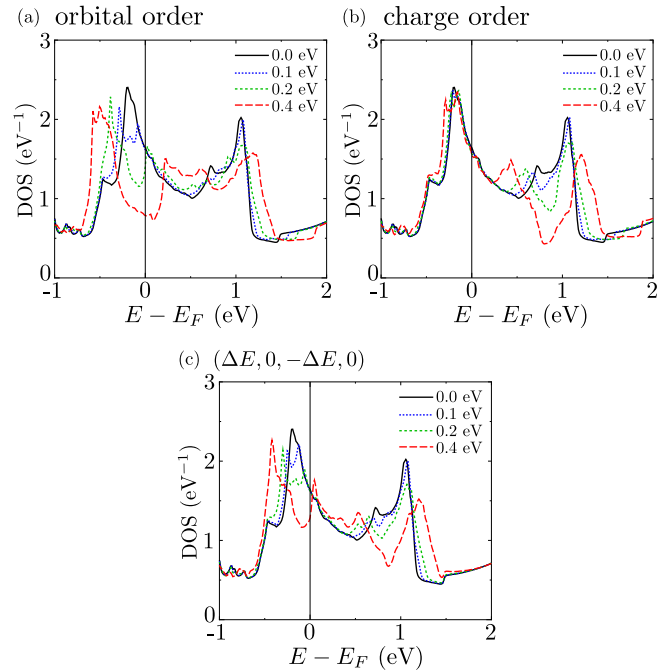}
\caption{
(color online)
The DOSs for (a) orbital order state $\Delta {\bm E}_{\rm orbital}$,
(b) charge order state $\Delta {\bm E}_{\rm charge}$, and 
(c) $\Delta {\bm E} \equiv (\Delta E,0,-\Delta E,0)$.
%=(\Delta {\bm E}_{\rm orbital}+\Delta {\bm E}_{\rm charge})/2$.
The black, blue, green and red lines correspond to 
$\Delta E=0$, $0.1$, $0.2$, and $0.4$ eV respectively.
}
\label{fig:DOS}
\end{figure}

Figure \ref{fig:DOS} shows the DOS, $D(\e)$, in the 
(a) orbital-ordered state and (b) charge-ordered state
for $\Delta E=0\sim0.4$ eV.
To make comparison with experiments qualitatively, $\Delta E$
should be multiplied with the renormalization factor 
due to the self-energy, $z\equiv m_{\rm band\mbox{-}calc}/m^* \ (<1)$, although 
the value of $z$ is unknown in Ti-oxypnictides.
(Note that $z^{-1}\approx 2\sim10$ in Fe-based superconductors.)
In (a), the DOS at the Fermi level, $D(0)$, decreases with $\Delta E >0$,
and the pseudo-gap structure appears.
In (b), in contrast, $D(0)$ is almost independent of $\Delta E$.
The reason for the pseudo-gap formation in (a) is that 
both the electron-like FS around M point and hole-like FS around X point
shrink with $\Delta E$,
whereas only the latter shrinks in (b).
The striking difference in the FS deformation 
%in the orbital order and charge order 
is understood from the orbital character of the 
electron-like FS, as we discuss in Appendix \ref{sec:AP-C}.

In Fig. \ref{fig:DOS} (c), we show the DOS for
$\Delta {\bm E} \equiv (\Delta E,0,-\Delta E,0)
[=(\Delta {\bm E}_{\rm orbital}+\Delta {\bm E}_{\rm charge})/2]$,
which is also a possible nematic state suggested experimentally
\cite{Frandsen2014}.

Experimentally, below $T_S$,
the resistivity shows the upturn, and the 
uniform susceptibility is suppressed \cite{YajimaJPSJ2012,YajimaJPSJ2013}.
%in various Ti-oxypnictides.
%The electronic specific heat $C$ shows a BCS-like jump
%at $T\approx T_S$, and largely suppressed for $T\ll T_S$.
Also, a pseudo-gap formation is indicated by ARPES below $T_S$
\cite{Tan,Xu}.
Thus, the reduction of $D(0)$
due to the orbital order shown in  Fig. \ref{fig:DOS} (a) 
is consistent with experimental results in Ti-oxypnictides.

%%%%%%%%%%%%%%%%%%%%%%%%%%%%%%%%%
%%%%%%%%%%%%%%%%%%%%%%%%%%%%%%%%%

%%%%%%%%%%%%%%%%%%
\section{Discussions}
\label{sec:discuss}
%%%%%%%%%%%%%%%%%%

We discuss the mechanism of the superconductivity
in Ti-oxypnictides.
Up to now, both the spin-fluctuation-mediated unconventional pairing 
\cite{Singh}
and the phonon-mediated conventional pairing
\cite{Subedi}
mechanisms were proposed.
The latter may be supported by the
full-gap structure reported by 
the specific heat study in Ba$_{1-x}$Na$_x$Ti$_2$Sb$_2$O
\cite{Gooch}.
However, it is naturally expected that the orbital fluctuations 
near the orthorhombic phase would contribute to the pairing mechanism
 \cite{Kontani-RPA}.
This is our important future problem.

In the present AL-VC theory,
the obtained wavevector of the orbital order is $\q=(0,0,0)$,
which is consistent with the report in Ref. \cite{Frandsen2014}.
On the other hand, the SC-VC theory explain the 
{\it incommensurate orbital order} at $\q_c=(\delta,0,0)$
in cuprate superconductors \cite{Yamakawa-CDW,Tsuchiizu-CDW}:
$\q_c$ is equal to the wavevector connecting the neighboring hot spots.
This fact indicates that the 
incommensurate orbital order might be realized in some Ti-oxypnictides,
depending on the details of the bandstructure.

In Ref. \cite{Frandsen2014}, the authors 
discussed the nematic charge order driven by the 
nearest-neighbor Coulomb interaction $V$.
At present, there is no first principles study for $V$.
However, if $V$ were the origin of the nematic order, 
fine tuning of the parameters $V$ and $U$ is required 
to explain the development of spin fluctuations near $T_S$
in BaTi$_2$Sb$_2$O \cite{NMR}.
On the other hand, the coexistence of spin and orbital fluctuations 
is naturally explained by the strong orbital-spin interplay
described by the AL-VC. 
The AL-VC mechanism explains the orbital-order
in Fe-based superconductors \cite{Onari-SCVC}
and the nematic CDW order in cuprate superconductors 
\cite{Yamakawa-CDW}.
We stress that the importance of the AL-VC has been confirmed by 
the unbiased numerical study 
using the functional-renormalization-group theory
\cite{Tsuchiizu-CDW,Tsuchiizu-RG}

In summary, 
we studied the origin of the nematic order without magnetization 
in Ti-oxypnictides based on the three-dimensional first-principles model. 
%for BaTi$_2$(As,Sb)$_2$O.
%Due to the FS nesting, the AFM fluctuations develop 
%at $\Q_s=(\pi,0,\pi)$ and $(0,\pi,\pi)$.
%The strong orbital fluctuations at $\q=(0,0,0)$
We predicted the formation of the intra-unit-cell orbital order in 
BaTi$_2$(As,Sb)$_2$O, which is driven by the orbital-spin interplay (AL-VC).
The present intra-unit-cell orbital order can be confirmed experimentally 
by observing the shear modulus $C_S$ and the electron Raman spectroscopy 
for $B_{1g}$ channel, both of which are useful to observe the 
nematic fluctuations in Fe-based superconductors
\cite{Kontani-Raman}. 
%Thus, the spin-fluctuation-driven orbital order due to the 
The orbital order due to the AL-VC mechanism 
is expected to emerge not only in 
the two-dimensional high-$T_{\rm c}$ superconductors, but also in 
the three-dimensional multiorbital systems with moderate spin fluctuations
such as Ti-oxypnictides.
It would be an interesting future problem to clarify
the role of the orbital fluctuations on the superconductivity.

%%%%%%%%%%%%%%%%%%%%%
\acknowledgements
We are grateful to T. Yajima for useful discussions.
This study has been supported by Grants-in-Aid for Scientific 
Research from MEXT of Japan.

%%%%%%%%%%%%%%%%%%%%%%%%%%%%%%%%
\appendix
\section{Numerical study of the AL-VC in three-dimensional systems}
\label{sec:AP-A}

In the main text, we study the Aslamazov-Larkin vertex correction (AL-VC)
in the three-dimensional 10-orbital Hubbard model for Ti-oxypnictides.
To our knowledge, this is the first numerical study of the 
AL-VC in the three-dimensional multiorbital model.

\begin{figure}[htbp]
\includegraphics[clip,width=6.5cm]{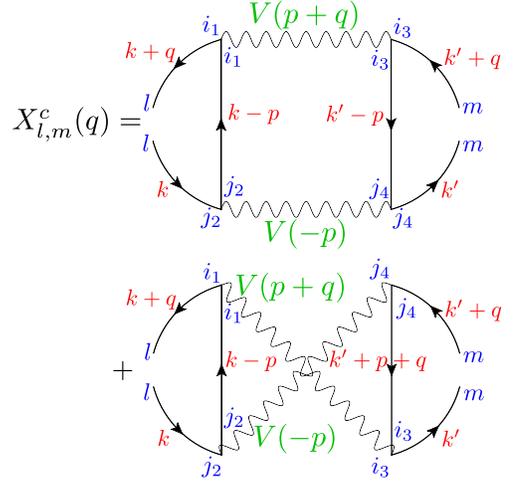}
\caption{
The diagram of AL-VC. $X^{c}_{l,m}(q)=X^{c}_{ll,mm}(q)$
}
\label{fig2}
\end{figure}

In this model, we verified that the AL-VCs $X^{c}_{l,l';m,m'}(q)$
with $l\ne l'$ or $m\ne m'$ are negligibly small.
Therefore, we calculate the AL-VC $X^{c}_{l,m}(q)\equiv X^{c}_{l,l;m,m}(q)$
in the present study.
Its expression is given as
\begin{eqnarray}
X^{c}_{l,m}(q)&=&\frac{T}{2}\sum_{p}\sum_{i_{1}\sim j_{4}}\Lambda_{ll,i_{1}i_{2},j_{1}j_{2}}(q;p)\Lambda'_{mm,i_{3}i_{4},j_{3}j_{4}}(q;p) \nonumber \\
&&\times\{3V^{s}_{i_{1}i_{2},i_{3}i_{4}}(p+q)V^{s}_{j_{1}j_{2},j_{3}j_{4}}(-p) \nonumber \\
&&\hspace{0.4cm}+V^{c}_{i_{1}i_{2},i_{3}i_{4}}(p+q)V^{c}_{j_{1}j_{2},j_{3}j_{4}}(-p)\} ,
\label{eqn:AL-exp}
\end{eqnarray}
where the three-point vertex and spin (charge) channel interaction are given as
\begin{eqnarray}
&&\!\!\!\!\!\!\!\! \Lambda_{ll,i_{1}i_{2},j_{1}j_{2}}(q;p)\nonumber \\
&& \ \ \ \ \ \ =-T\sum_{k}G_{li_{1}}(k+q)G_{j_{2}l}(k)G_{i_{2}j_{1}}(k-p) , \\
&&\!\!\!\!\!\!\!\! \Lambda'_{mm,i_{3}i_{4},j_{3}j_{4}}(q;p)
=\Lambda_{i_{3}j_{4},mj_{3},i_{4}m}(q;p) \nonumber \\
&&\ \ \ \ \ \ \ \ \ \ \ \ \ \ \ \ \ \ \ \ \ \ 
+\Lambda_{j_{3}i_{4},mi_{3},j_{4}m}(q;-p-q) , \\
&&\!\!\!\!\!\!\!\! {\hat V}^{s(c)}(q)
={\hat \Gamma}^{s(c)}+{\hat \Gamma}^{s(c)}{\hat \chi}^{s(c)}(q){\hat \Gamma}^{s(c)} .
\end{eqnarray}
The diagrammatic expression of the AL-VC is shown in Fig. \ref{fig2}.
Here, the second-order double counting terms in the AL-VC should be subtracted.
In the present numerical study, 
we put $i_{1}=i_{2}$, $i_{3}=i_{4}$, $j_{1}=j_{2}$ and $j_{3}=j_{4}$ 
in eq. (\ref{eqn:AL-exp}).
This simplification is justified since $\chi^{s}_{l,l';m,m'}(q)$
is negligibly small for $l\ne l'$ or $m\ne m'$.

%%%%%%%%%%%%%%%%%%%%%%%%%%%%%%%%%%%%%%%%%
%%%%%%%%%%%%%%%%%%%%%%%%%%%%%%%%%%%%%%%%%

In general, the AL-VC is less important in higher-dimensional systems,
so it is significant to verify whether orbital fluctuations 
due to the AL-VC could develop in three-dimensional systems.
In the presence of strong spin fluctuations, the AL-VC is scaled as 
\begin{eqnarray}
X^{c}({\bm 0})\sim |\Lambda({\bm 0};\Q_{s})|^{2}T\sum_{p}\{V^{s}(\p,0)\}^2,
\end{eqnarray} 
which is proportional to $|\Lambda({\bm 0};\Q_{s})|^{2}T\xi_{\rm AF}^{4-d}$
when the spin fluctuations are $d$-dimensional,
where $\xi_{\rm AF}$ is the antiferromagnetic correlation length.
Therefore, the AL-VC is expected to be less important 
in the three-dimensional systems.
Nonetheless, as revealed by the numerical study in the main text,
the strong orbital fluctuations due to the AL-VC is realized 
in Ti-oxypnictides with three-dimensional FSs,

 %%%%%%%%%%%%%%%%%%%%%%%%%%%%
%\subsection{Why intra-unit-cell orbital order is induced by the AL-VC?}
\section{Mathematical explanation for the
intra-unit-cell orbital order due to the AL-VC}
\label{sec:AP-B}

In the main text, we obtained the ``intra-unit-cell orbital order'' 
based on the BaTi$_2$(As,Sb)$_2$O model, 
by taking the AL-VC for ${\hat \chi}^c(\q)$ into account.
The main origin of the orbital order is the large diagonal elements 
of the AL-VC $X_{l,l;l,l}(\q)$ at $\q={\bm 0}$,
which is enlarged near the magnetic critical point.
%$\xi_{\rm AF}^2\propto (T-\theta_S)^{-1}$.
Each $\chi_{l,l;m,m}^c({\bm 0})$ ($l,m=1\sim4$)
has positive or negative large value,
%as (partially) shown in Figs. \ref{fig:SCVC} (a)-(c),
and the obtained orbital and charge susceptibilities are shown in
Figs. \ref{fig:SCVC} (a) and (c).

However, it is nontrivial that inter-site elements of  
$|\chi_{l,l;m,m}^c({\bm 0})|$ with $l\le2$ and $m\ge3$ are enlarged by 
the diagonal elements of the AL-VC and intra-site Coulomb interaction.
We present the explanation for this nontrivial question 
based on the equation for the susceptibility (\ref{eqn:chi}).
\begin{table}[h]
\centering
\begin{tabular}{|r||r|r|r|r|}
\hline
$\chi^0_{l,l;m,m}$ & $m=1$ & $2$ & $3$ & $4$ \\
\hline
\hline
$l=1$ & $\ \bm{0.364}$ & $-0.037$ & $-0.034$ & $0.094$ \\
\hline
$2$ & & $\bm{0.258}$ & $0.094$ & $-0.046$  \\
\hline
$3$ &  &  & $\bm{0.364}$ & $-0.037$ \\
\hline
$4$ &  &  &  & $\bm{0.258}$ \\
\hline
\end{tabular}

\vspace{3mm}

\begin{tabular}{|r||r|r|r|r|}
\hline
$X^{c}_{l,l;m,m}$ & $m=1$ & $2$ & $3$ & $4$ \\
\hline
\hline
$l=1$ & $\ \bm{0.844}$ & $\ 0.020$ & $\bm{-0.252}$ & \ $0.013$ \\
\hline
$2$ & & $\bm{0.487}$ & $0.013$ & $0.099$  \\
\hline
$3$ &  &  & $\bm{0.844}$ & $0.020$ \\
\hline
$4$ &  &  &  & $\bm{0.487}$ \\
\hline
\end{tabular}
\caption{\label{tab:X}
Bare susceptibility $\chi_{l,l;m,m}^0$ (upper table) 
and AL-VC for the charge channel $X_{l,l;m,m}^c$ (lower table) at $\q={\bf 0}$
in the case of $U=2.06$ eV and $J/U=1/9$ at $T=50$ meV.
The Stoner factors are $(\a_C,\a_S)=(0.99,0.98)$.
The numbers greater than 0.2 in magnitude are shown by bold fonts.
}
\end{table}
Table \ref{tab:X} shows the lists of the
bare susceptibility $\chi_{l,l;m,m}^0$ (upper table) 
and AL-VC for the charge channel $X_{l,l;m,m}^c$ (lower table) at $\q={\bm 0}$.
The model parameters are equal to those used in the main text.
It is found that the irreducible susceptibility 
$\Phi_{l,l;m,m}^c=\chi_{l,l;m,m}^0+X_{l,l;m,m}^c$
takes large value in magnitude only for $l=m=1\sim4$ and for $(l,m)=(1,3)$. 
In addition, the elements $\Phi_{l,l';m,m'}$ with $l\ne l'$ or $m\ne m'$
are very small.
In this case, the simplified equation for 
$\chi^c_{l,l;m,m}$, given by the $4\times4$ matrix, is expressed as
\begin{eqnarray}
{\hat {\chi}}^{c(s)}
= (1-{\hat \Phi}^{c(s)}{\hat{\Gamma}}^{c(s)})^{-1}{\hat \Phi}^{c(s)} .
\label{eqn:chi-AP}
\end{eqnarray}
In this section,
we denote ${\chi}^{c(s)}_{l,m}\equiv \chi^{c(s)}_{l,l;m,m}$ and
$\Phi_{l,m}^{c(s)}\equiv \Phi_{l,l;m,m}^{c(s)}$, and 
${\Gamma}^{c(s)}_{l,m}\equiv \Gamma^{c(s)}_{l,l;m,m}$.

First, we analyze the intra-site charge susceptibility with $l,m\le2$
by neglecting the inter-site $\Phi_{l,m}^c$ ($l\le2$, $m\ge3$).
The $2\times2$ charge susceptibility is given as
\begin{align}
\begin{pmatrix}
{\chi}_{1,1}^c  & {\chi}_{1,2}^c \\
{\chi}_{2,1}^c & {\chi}_{2,2}^c \hfill 
\end{pmatrix}
&=
({\hat 1}-{\hat C}_2)^{-1}
\begin{pmatrix}
\Phi_{1,1}^c & \Phi_{1,2}^c \\
\Phi_{2,1}^c & \Phi_{2,2}^c 
\end{pmatrix} ,
\\
{\hat C}_2
&=
\begin{pmatrix}
\Phi_{1,1}^c & \Phi_{1,2}^c \\
\Phi_{2,1}^c & \Phi_{2,2}^c 
\end{pmatrix} 
\!\!
\begin{pmatrix}
-U & -2U'+J \\
-2U'+J & -U 
\end{pmatrix} .
\end{align}
The maximum eigenvalue of $\displaystyle {\hat C}_2$
gives the charge Stoner factor $\a_C$, and its
eigenvector gives the form factor ${\bm g}$ at $\a_C=1$.
(Note that $\displaystyle {\hat C}_2$ is not Hermitian.)
For simplicity, we examine the case of $
\Phi_{l,m}^c=0$ for $l\ne m$, and $U'=U$ and $J=0$.
In this case, the charge Stoner factor is
$\a_C= \frac U2 [-(\Phi_{1,1}^c+\Phi_{2,2}^c)+
\sqrt{{\Phi_{1,1}^c}^2+{\Phi_{2,2}^c}^2+14{\Phi_{1,1}^c}{\Phi_{2,2}^c}}]\ (>0)$,
and the form factor is
${\bm g}\propto (-{\Phi_{1,1}^c}+{\Phi_{2,2}^c}+
\sqrt{{\Phi_{1,1}^c}^2+{\Phi_{2,2}^c}^2+14{\Phi_{1,1}^c}{\Phi_{2,2}^c}}, \
-4{\Phi_{2,2}^c})$.
When $\Phi_{2,2}=(1-x)\Phi_{1,1}$ and $|x|\ll1$,
the form factor is simplified as 
${\bm g}\propto (y,-1)$,
where $y=(\sqrt{1-x+x^2/16} -x/4)/(1-x) \approx 1+x/4$.

For the $4\times4$ charge susceptibility in Eq. (\ref{eqn:chi-AP}),
the charge Stoner factor and the form factor 
are respectively given by the maximum eigenvalue and its eigenvector 
of the following $4\times4$ matrix:
\begin{align}
{\hat C}_4
&= 
\begin{pmatrix}
{\hat C}_2 & {\hat C}_2' \\
{\hat C}_2'' & {\hat C}_2 
\end{pmatrix} ,
 \\
{\hat C}_2'
&=
\begin{pmatrix}
\Phi_{1,3}^c & \Phi_{1,4}^c \\
\Phi_{2,3}^c & \Phi_{2,4}^c 
\end{pmatrix} 
\!\!
\begin{pmatrix}
-U & -2U'+J \\
-2U'+J & -U 
\end{pmatrix} ,
\\
{\hat C}_2''
&=
\begin{pmatrix}
\Phi_{3,1}^c & \Phi_{4,1}^c \\
\Phi_{3,2}^c & \Phi_{4,2}^c 
\end{pmatrix} 
\!\!
\begin{pmatrix}
-U & -2U'+J \\
-2U'+J & -U 
\end{pmatrix} .
\end{align}
At $\q=0$, the relation ${\hat C}_2'= {\hat C}_2''$ holds, and
the form factors are $({\bm g},\pm{\bm g})$ when ${\hat C}_2'=0$.
This degeneracy of the form factor
is lifted in the presence of small ${\hat C}_2'$,
and the form factor is given as
${\bm f}\approx({\bm g},-{\bm g})$ when the inner product
$({\bm g},{\hat C}_2'{\bm g})$ is negative,
which is satisfied in the present numerical study
due to the large negative $X_{1,1;3,3}^c({\bm 0})$ in Table \ref{tab:X}.

To summarize, we presented a mathematical explanation why the 
orbital-order with the form factor ${\bm f}\sim (1,-1,-1,1)$,
which corresponds to the numerical results in 
Figs. \ref{fig:SCVC} (a) and (c),
is universally obtained in the present numerical study.
%for BaTi$_2$(As,Sb)$_2$O.

%In the same way, 
Next,
we analyze the intra-site spin susceptibility with $l,m\le2$
by neglecting the inter-site $\Phi_{l,m}^s$ ($l\le2$, $m\ge3$).
The $2\times2$ spin susceptibility is given as
\begin{align}
\begin{pmatrix}
{\chi}_{1,1}^s & {\chi}_{1,2}^s \\
{\chi}_{2,1}^s & {\chi}_{2,2}^s \hfill 
\end{pmatrix}
&=
({\hat 1}-{\hat S}_2)^{-1}
\begin{pmatrix}
\Phi_{1,1}^s & \Phi_{1,2}^s \\
\Phi_{2,1}^s & \Phi_{2,2}^s 
\end{pmatrix} ,
\\
{\hat S}_2&=
\begin{pmatrix}
\Phi_{1,1}^s & \Phi_{1,2}^s \\
\Phi_{2,1}^s & \Phi_{2,2}^s 
\end{pmatrix} 
\!\!
\begin{pmatrix}
U & J \\
J & U 
\end{pmatrix} ,
\end{align}
where $\Phi_{l,m}^s(\q)\approx \chi_{l,m}^0({\bm q})$
is satisfied since the AL-VC for the spin channel is small.
Since $J\ll U$,
only the orbital diagonal elements of the spin susceptibility 
${\chi}_{l,m}^s$ are enlarged.
The spin Stoner factor is obtained as
$\a_S=\frac U2 [ (\Phi_{1,1}^s+\Phi_{2,2}^s)
+\sqrt{(\Phi_{1,1}^s-\Phi_{2,2}^s)^2 + 4(J/U)^2\Phi_{1,1}^s\Phi_{2,2}^s} ]$
when $\Phi_{l,m}^s$ with $l\ne m$ is negligible.

In the RPA or FLEX approximation (${\hat \Phi}^s={\hat \Phi}^c$),
the relation $\a_C\le\a_S$ holds at any $\q$ for $J\ge0$,
according to the obtained expressions for $\a_C$ and $\a_S$.
%Therefore, the magnetization occurs first within the RPA.
Nonetheless, the opposite relation $\a_C>\a_S$ is obtained 
in the present study thanks to the charge channel AL-VC.
Therefore, the ``orbital order without magnetization'' 
is explained by the present orbital-spin fluctuation theory
with including the AL-VC.

%%%%%%%%%%%%%%%%%%%%%%%%%%%%%
\section{Fermi surface deformation due to intra-unit-cell orders}
\label{sec:AP-C}

%%%%%%%%%%%%%%%%%%%%%%%%%%%%%%%%%%%%%%%%
\begin{figure}[htbp]
\includegraphics[width=0.9\linewidth]{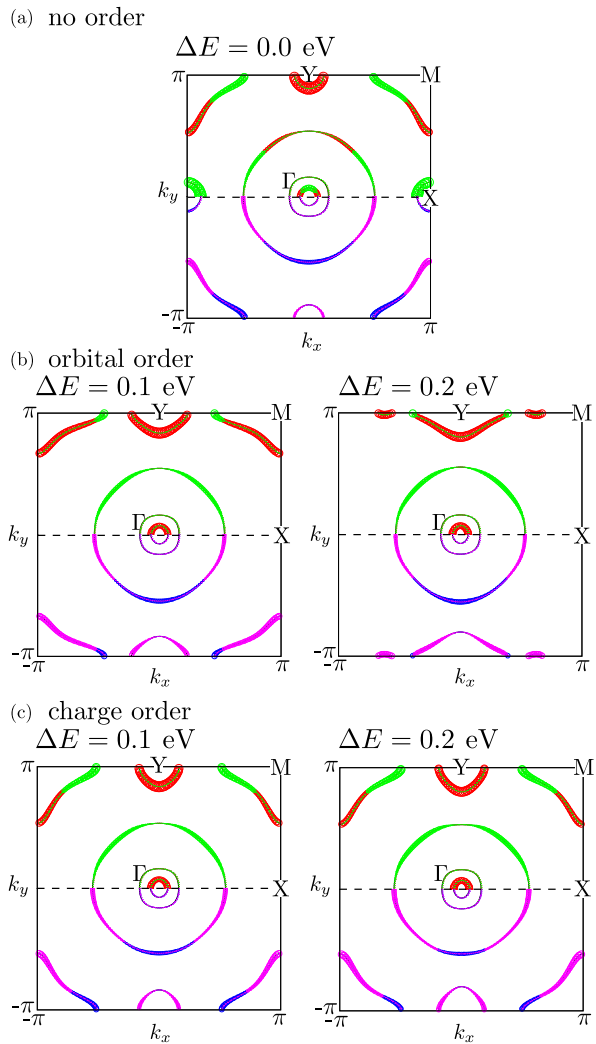}
\caption{
(a) The FSs in the $k_z=0$ plane in the absence of the orbital or charge order.
The weights of the 1-, 2-, 3-,  4-orbitals are 
shown by the red, blue, green, and purple  circles, respectively.
The weights of the 1,3-orbitals (2,4-orbitals) are shown 
in the upper (lower) Brillouin zone.
(b) The FSs under the orbital order with $\Delta E=0.1$ eV and $0.2$ eV.
(c) The FSs under the charge order with $\Delta E=0.1$ eV and $0.2$ eV.
}
\label{fig:FS-orbital}
\end{figure}
%%%%%%%%%%%%%%%%%%%%%%%%%%%%%%%%%%%%%%%%

Here, we examine the orbital character of the FSs in detail,
and discuss the FS deformation under the orbital and charge orders.
Figure \ref{fig:FS-orbital} (a) 
shows the FSs of the original tight-binding model
in the $k_z=0$ plane.
The weights of the 1,3-orbitals (2,4-orbitals) are shown 
in the upper (lower) Brillouin zone.

First, we calculate the FS deformation under the intra-unit-cell
orbital order, by introducing the potential
$\Delta {\bm E}_{\rm orbital} \equiv (\Delta E,-\Delta E,-\Delta E,\Delta E)$.
%$\Delta {\bm E}_{\rm orbital} \equiv (1,-1,-1,1)\times\Delta E$.
%,-\Delta E,-\Delta E,\Delta E)$.
Figure \ref{fig:FS-orbital} (b) shows the FSs for $\Delta E=0.1$ eV 
and $0.2$ eV.
In this case, the hole-FS around the X-point, which is 
mainly composed of the 3-orbital (green),
disappears for $\Delta E\ge0.1$ eV.
In addition, the electron-FS around the M-point,
mainly composed by (1+4)-orbital [(2+3)-orbital] 
near the $k_x=\pi$ [$k_y=\pi$] Brillouin zone boundary,
disappears for $\Delta E\ge0.2$ eV.
For this reason, the pseudo-gap structure appears in the DOS
at the Fermi level, as shown in Fig. 5 (a) in the main text.

Next, we calculate the FS deformation under the intra-unit-cell charge order 
with $\Delta {\bm E}_{\rm charge} \equiv (\Delta E,\Delta E,-\Delta E,-\Delta E)$.
%$\Delta {\bm E}_{\rm charge} \equiv (1,1,-1,-1)\times\Delta E$.
%$\Delta {\bm E}' \equiv (\Delta E,\Delta E,-\Delta E,-\Delta E)$.
Figure \ref{fig:FS-orbital} (c) 
shows the FSs for $\Delta E=0.1$ eV and $0.2$ eV.
Then, the hole-FS around the X-point
disappears for $\Delta E\ge0.1$ eV, similarly to Fig. \ref{fig:FS-orbital} (b).
However, in contrast, the electron-FS around the M-point
still exists for $\Delta E=0.2$ eV,
since the potential on the (1+4)-orbital and that on the (2+3)-orbital
cancel in the case of the charge order.
For this reason, the DOS at the Fermi level is insensitive to $\Delta E$,
as shown in Fig. 5 (b) in the main text.

According to the ARPES study for Na$_{2}$Ti$_{2}$Sb$_{2}$O, 
the pseudo-gap appears around the X-point~\cite{Tan}. 
This result is consistent with the FS deformation due to orbital order
in Fig. \ref{fig:FS-orbital} (b) as well as that due to the charge order 
in Fig. \ref{fig:FS-orbital} (c).
To distinguish between the orbital order and charge order,
the ARPES study around the M-point is required.

%%%%%%%%%%%%%%%%%%%%%%%%
%references
%%%%%%%%%%%%%%%%%%%%%%%%

\end{document}